\documentstyle[12pt]{article}
\renewcommand{\baselinestretch}{1.2}

\begin{document}

\begin{flushright}
{\bf hep-ph/9907454} \\
{\bf LMU-99-11} \\
July 1999
\end{flushright}

\vspace{0.2cm}

\begin{center}
{\large\bf New Time Distributions of $D^0$-$\bar{D}^0$ or
$B^0$-$\bar{B}^0$ Mixing and $CP$ Violation}
\end{center}

\vspace{.5cm}
\begin{center}
{\bf Zhi-zhong Xing} \footnote{
E-mail: Xing$@$hep.physik.uni-muenchen.de }\\
{\sl Sektion Physik, Universit${\sl\ddot a}$t M${\sl\ddot u}$nchen, 
Theresienstrasse 37A, 80333 M${\sl\ddot u}$nchen, Germany}
\end{center}

\vspace{3.5cm}

\begin{abstract}
The formulae for $D^0$-$\bar{D}^0$ or $B^0$-$\bar{B}^0$ mixing 
and $CP$ violation at the $\tau$-charm or $B$-meson factories
are derived, for the
case that only the decay-time distribution of one $D$ or $B$ meson is
to be measured. In particular, we point out 
a new possibility to determine the $D^0$-$\bar{D}^0$ mixing rate
in semileptonic $D$ decays at the $\Psi (4.14)$ resonance;
and show that both direct
and indirect $CP$ asymmetries can be measured at the $\Upsilon (4S)$ 
resonance without ordering the decay times of two $B_d$ mesons or 
measuring their difference.
\end{abstract}

\newpage

{\Large\bf 1} ~
It is well known that mixing between a neutral
meson and its $CP$-conjugate counterpart can arise if both of them
couple to a subset of real and (or) virtual intermediate states.
Such mixing effects provide a mechanism whereby interference between
the decay amplitudes of two mesons may occur, leading to the
phenomenon of $CP$ violation. To date the $K^0$-$\bar{K}^0$ 
and $B^0_d$-$\bar{B}^0_d$ mixing rates have been measured \cite{PDG}, 
and the $CP$-violating signals in neutral $K$-meson decays have
unambiguously been established \cite{CP}. A preliminary but encouraging
result for the observation of $CP$ violation in 
$B^0_d$ vs $\bar{B}^0_d \rightarrow J/\psi K_{\rm S}$
decay modes has recently been reported by the CDF Collaboration \cite{CDF}.
In contrast, the present experiments have only yielded
the upper bound on $D^0$-$\bar{D}^0$ mixing 
and the lower bound on $B^0_s$-$\bar{B}^0_s$ mixing \cite{PDG}, which
are respectively expected to be rather small and large in
the standard model. Today the $B^0_d$-$\bar{B}^0_d$ and
$B^0_s$-$\bar{B}^0_s$ 
systems are playing important roles in the study of flavor
mixing and $CP$ violation beyond the neutral kaon system.
The $D^0$-$\bar{D}^0$ system is, on the other hand, 
of particular interest to probe possible new physics 
that might give rise to observable $D^0$-$\bar{D}^0$ mixing
and $CP$ violation in the charm sector.

\vspace{0.4cm}

The most promising place to produce $B^0_d$ and $\bar{B}^0_d$ events 
with high statistics and low backgrounds is the 
$\Upsilon (4S)$ resonance, on which the asymmetric $B$-meson
factories at KEK and SLAC as well as the symmetric 
$B$-meson factory at Cornell are based. Similarly 
$B^0_s$ and $\bar{B}^0_s$ events may coherently be produced 
at the $\Upsilon (5S)$ resonance. At a $\tau$-charm factory
$D^0$ and $\bar{D}^0$ events will in huge amounts be produced
at the $\Psi (4.14)$ resonance. To measure $CP$ violation on
any resonance, where the produced meson pair has the odd 
charge-conjugation parity ($C=-1$), 
a determination of the time interval between 
two meson decays is generally needed. This has led
to the idea of asymmetric $e^+e^-$ collisions at the 
$\Upsilon (4S)$ resonance, i.e., asymmetric $B$-meson
factories, in which the large boost allows to order the decay times
of two $B_d$ mesons and to measure their difference.

\vspace{0.4cm}

Recently a new idea, that $CP$ violation can be measured
on the $\Upsilon (4S)$ resonance without ordering the decay times
of two $B_d$ mesons or determining their difference, has been
pointed out by Foland \cite{Foland}. If this idea is
really feasible, it implies that the 
time-dependent measurement of $B^0_d$-$\bar{B}^0_d$ mixing and
$CP$ violation may be realized at a symmetric $e^+e^-$ collider
running at the $\Upsilon (4S)$ resonance, such as the one
operated by the CLEO Collaboration at Cornell.
It also implies that the time-dependent measurement 
of $D^0$-$\bar{D}^0$ mixing and $CP$ violation may straightforwardly
be carried out at the $\Psi (4.14)$ resonance with no need
to build an asymmetric $\tau$-charm factory. Therefore 
a further and more extensive exploration of Foland's idea and
its consequences is desirable.

\vspace{0.4cm}

This note aims at reformulating the phenomenology of 
meson-antimeson mixing and $CP$ violation at the $\Upsilon (4S)$,
$\Upsilon (5S)$ or $\Psi (4.14)$ 
resonance, for the case that only the decay-time distribution
of one meson is to be measured. We take both $C=-1$ and $C=+1$ cases of
the produced meson pair into account, and make no special
assumption in deriving the generic formulae.
In particular, we point out 
a new possibility to determine the $D^0$-$\bar{D}^0$ mixing rate
in semileptonic $D$ decays at the $\Psi (4.14)$ resonance;
and show that both direct
and indirect $CP$ asymmetries can be measured at the $\Upsilon (4S)$ 
resonance without ordering the decay times of two $B_d$ mesons or 
measuring their difference.

\vspace{0.4cm}

{\Large\bf 2} ~
Let us make use of $P$ to symbolically denote $D$, $B_d$ or $B_s$ meson.
In the assumption of $CPT$ invariance, the mass eigenstates of $P^0$ 
and $\bar{P}^0$ mesons can be written as
\begin{eqnarray}
|P_{\rm L}\rangle & = & p |P^0\rangle ~ + ~ q |\bar{P}^0\rangle \; , \nonumber \\
|P_{\rm H}\rangle & = & p |P^0\rangle ~ - ~ q |\bar{P}^0\rangle \; ,
%		(1)
\end{eqnarray}
in which the subscripts ``L'' and ``H'' stand for Light and Heavy respectively,
and ($p, q$) are complex mixing parameters. 
The proper-time evolution of an initially ($t=0$) pure 
$P^0$ or $\bar{P}^0$ meson is given as
\begin{eqnarray}
|P^0 (t)\rangle & = & g^{~}_+(t) |P^0\rangle ~ 
+ ~ \frac{q}{p} g^{~}_-(t) |\bar{P}^0\rangle \; , \nonumber \\
|\bar{P}^0 (t)\rangle & = & g^{~}_{+}(t)|\bar{P}^0\rangle ~ 
+ ~ \frac{p}{q} g^{~}_-(t) |P^0\rangle \; ,
%		(2)
\end{eqnarray}
where
\begin{eqnarray}
g^{~}_+(t) & = & \exp \left [ -\left ({\rm i} m +\frac{\Gamma}{2} \right ) t \right ] 
\cosh \left [ \left ( {\rm i} \Delta m - \frac{\Delta\Gamma}{2} \right ) \frac{t}{2} \right ] \; , \nonumber \\
g^{~}_-(t) & = & \exp \left [ -\left ({\rm i} m + \frac{\Gamma}{2} \right ) t \right ]
\sinh \left [ \left ( {\rm i} \Delta m - \frac{\Delta\Gamma}{2} \right ) \frac{t}{2} \right ] \; , 
%		(3)
\end{eqnarray}
with the definitions
$m = (m^{~}_{\rm L}+m^{~}_{\rm H})/2$,
$\Delta m = m^{~}_{\rm H} - m^{~}_{\rm L}$,
$\Gamma = (\Gamma_{\rm L}+\Gamma_{\rm H})/2$, and
$\Delta\Gamma = \Gamma_{\rm L} - \Gamma_{\rm H}$.
Here $m^{~}_{\rm L(H)}$ and $\Gamma_{\rm L(H)}$ are the mass and width 
of $P_{\rm L(H)}$, respectively.
In practice it is more popular to use two dimensionless 
parameters for the description of $P^0$-$\bar{P}^0$ mixing:
$x = \Delta m/\Gamma$ and
$y = \Delta\Gamma/(2\Gamma)$.

\vspace{0.4cm}

For a coherent $P^0 \bar{P}^0$ pair at rest, 
its time-dependent wave function can be written as
\begin{equation}
\frac{1}{\sqrt{2}} \left [| P^0 ({\bf K},t)\rangle \otimes 
|\bar{P}^0 (-{\bf K}, t)\rangle
~ + ~ C |P^0 (-{\bf K}, t)\rangle \otimes 
|\bar{P}^0 ({\bf K}, t)\rangle \right ] \;  ,
%		(4)
\end{equation}
where $\bf K$ is the three-momentum vector of the $P$ mesons, 
and $C=\pm 1$ denotes the
charge-conjugation parity of this coherent system. 
The formulae for the time evolution of
$P^0$ and $\bar{P}^0$ mesons have been given in Eq. (2). 
Here we consider the case that one of the two $P$ mesons 
(with momentum $\bf K$) decays to a final state
$f_1$ at proper time $t_1$ and the other (with $-\bf K$) to 
$f_2$ at $t_2$. $f_1$ and
$f_2$ may be either hadronic or semileptonic states. The amplitude of such a
joint decay mode is given by
\begin{eqnarray}
A(f_1, t_1; f_2, t_2)_C & = & \frac{1}{\sqrt{2}} A_{f_1}A_{f_2} \xi_C
\left [ g^{~}_+(t_1)g^{~}_-(t_2) + C g^{~}_-(t_1)g^{~}_+(t_2) \right ] 
+ \; \nonumber \\
& & \frac{1}{\sqrt{2}} A_{f_1}A_{f_2} \zeta_C 
\left [ g^{~}_+(t_1)g^{~}_+(t_2) + C g^{~}_-(t_1)g^{~}_-(t_2) \right ] \; ,
%		(5)
\end{eqnarray}
where $A_{f_i} = \langle f_i|{\cal H}|P^0\rangle$,
$\lambda_i = (q/p) (\langle f_i|{\cal H}|\bar{P}^0\rangle
/\langle f_i|{\cal H}|P^0\rangle)$ (for $i=1,2$), and
\begin{eqnarray}
\xi_C & = & \frac{p}{q} \left ( 1 ~ + ~ C \lambda^{~}_{f_1} 
\lambda^{~}_{f_2} \right ) \; , \nonumber \\
\zeta_C & = & \frac{p}{q} \left (\lambda^{~}_{f_2} ~ + ~ C 
\lambda^{~}_{f_1} \right ) \; .
%		(6)
\end{eqnarray}
After a lengthy calculation \cite{Xing96,Xing97}, we obtain the time-dependent 
decay rate as follows:
\begin{eqnarray}
R(f_1, t_1; f_2, t_2)_C & \propto & |A_{f_1}|^2|A_{f_2}|^2 
\exp (-\Gamma t_+) ~ \times  \nonumber \\
&  & \left [ \left (|\xi_C|^2 + |\zeta_C|^2 \right ) \cosh (y\Gamma t_C) 
-2 {\rm Re}\left (\xi^*_C \zeta_C \right ) \sinh (y\Gamma t_C) \right . 
\nonumber \\ 
&  & \left . - \left (|\xi_C|^2 - |\zeta_C|^2 \right ) \cos (x\Gamma t_C) 
+ 2{\rm Im}\left (\xi^*_C \zeta_C\right ) \sin (x\Gamma t_C) \right ] \; ,
%		(7)
\end{eqnarray}
where $t_C = t_2 + C t_1$ has been defined.

\vspace{0.4cm}

Now we integrate the decay rate $R(f_1, t_1; f_2, t_2)$ over 
$t_1 \in [0, \infty )$, i.e., only the time distribution
of $P$-meson decays into the final state $f_2$ is kept \cite{Foland}. 
The result, with the notation $t_2 =t$, is given as
\begin{eqnarray}
R(f_{1},f_{2}; t)_C & \propto & |A_{f_{1}}|^{2}|A_{f_{2}}|^{2}
\exp(-\Gamma t) ~ \times 
\nonumber \\
&  & \left [ \frac{ |\xi_C|^2 + |\zeta_C|^{2} }{\sqrt{1-y^2}}
\cosh(y \Gamma t + C\phi_y) - \frac{ 2{\rm Re} \left (\xi^*_C\zeta_C\right ) }
{\sqrt{1-y^2}} \sinh(y \Gamma t + C\phi_y) \right . 
\nonumber \\
& & \left . - \frac{ |\xi_C|^{2}-|\zeta_C|^{2} }{\sqrt{1+x^{2}}}
\cos \left (x \Gamma t + C\phi_{x}\right ) + 
\frac{ 2{\rm Im}\left (\xi^{*}_C\zeta_C\right ) }
{\sqrt{1+x^2}} \sin (x \Gamma t + C\phi_{x}) \right ]  \; ,
%		(8)
\end{eqnarray}
where the phase shifts $\phi_x$ and $\phi_y$ are defined by
$\tan\phi_{x}= x$ and $\tanh\phi_y = y$, respectively. 

\vspace{0.4cm}

The joint decay rate obtained above is a new result and
serves as the master formula
of this paper. In the following we shall specifically
investigate meson-antimeson
mixing and $CP$ violation in $D$- and $B$-meson decays into
the semileptonic final states, the hadronic $CP$ eigenstates,
and the hadronic non-$CP$ eigenstates.

\vspace{0.4cm}

{\Large\bf 3} ~
Let us first consider the joint decays of $(P^0\bar{P}^0)_C$
pairs into two semileptonic states $(l^{\pm} X^{\mp}_a)$ and
$(l^{\pm} X^{\mp}_b)$, i.e., the dilepton events in the final states.
Keeping the validity of the $\Delta Q = \Delta P$ rule and $CPT$
invariance, we have
$|\langle l^-X^+_i|{\cal H}|P^0\rangle| =
|\langle l^+X^-_i|{\cal H}|\bar{P}^0\rangle| =0$
and
$|\langle l^+X^-_i|{\cal H}|P^0\rangle| = 
|\langle l^- X^+_i|{\cal H}|\bar{P}^0\rangle| \neq 0$. The latter
is denoted later by $|A_{li}|$ for $i=a$ or $b$. With the help of Eq. (8),
we arrive at the same-sign and opposite-sign dilepton rates 
as follows:
\begin{eqnarray}
N^{++}_C(t) & \propto & \left |\frac{p}{q}\right |^2 
|A_{la}|^2 |A_{lb}|^2 \exp (-\Gamma t) \left [
\frac{\cosh (y\Gamma t + C\phi_y)}{\sqrt{1-y^2}} ~ - ~
\frac{\cos (x\Gamma t + C\phi_x)}{\sqrt{1+x^2}} \right ] \; ,
\nonumber \\
N^{--}_C(t) & \propto & \left |\frac{q}{p}\right |^2 
|A_{la}|^2 |A_{lb}|^2 \exp (-\Gamma t) \left [
\frac{\cosh (y\Gamma t + C\phi_y)}{\sqrt{1-y^2}} ~ - ~
\frac{\cos (x\Gamma t + C\phi_x)}{\sqrt{1+x^2}} \right ] \; ;
%		(9)
\end{eqnarray}
and
\begin{equation}
N^{+-}_C(t) \; \propto \; 2 |A_{la}|^2 |A_{lb}|^2 \exp (-\Gamma t) \left [
\frac{\cosh (y\Gamma t + C\phi_y)}{\sqrt{1-y^2}} ~ + ~
\frac{\cos (x\Gamma t + C\phi_x)}{\sqrt{1+x^2}} \right ] \; .
%		(10)
\end{equation}
Obviously the relationship 
$N^{++}_{+1}(t) N^{--}_{-1}(t) = N^{++}_{-1}(t) N^{--}_{+1}(t)$ holds.

\vspace{0.4cm}

The measure of $CP$ violation in $P^0$-$\bar{P}^0$ mixing 
turns out to be
\begin{equation}
{\cal A}^{+-}_C(t) \; =\; \frac{N^{++}_C(t) - N^{--}_C(t)}
{N^{++}_C(t) + N^{--}_C(t)} 
\; =\; \frac{|p|^4 - |q|^4}{|p|^4 + |q|^4} \; \; ,
%		(11)
\end{equation}
independent of both the decay time $t$ and the charge-conjugation
parity $C$. Within the standard model the magnitude of
${\cal A}^{+-}_C(t)$ is estimated to be of ${\cal O}(10^{-3})$ or
smaller, for either the $D^0$-$\bar{D}^0$ system \cite{Xing97} or the
$B^0$-$\bar{B}^0$ system \cite{Lusignoli,XingS97}. 
But it might significantly be enhanced if there were new physics
contributions to $P^0$-$\bar{P}^0$ mixing [6 -- 9].

\vspace{0.4cm}

On the other hand, the rate of $P^0$-$\bar{P}^0$ mixing 
can be determined from 
\begin{eqnarray}
S^{+-}_C(t) & = & \frac{N^{++}_C(t) + N^{--}_C(t)}{N^{+-}_C(t)} 
\nonumber \\
& = & \frac{1}{2} \left ( \left | \frac{p}{q} \right |^2 +
\left | \frac{q}{p} \right |^2 \right ) 
\frac{\cosh (y\Gamma t +C\phi_y) - z \cos (x\Gamma t +C\phi_x)}
{\cosh (y\Gamma t +C\phi_y) + z \cos (x\Gamma t + C\phi_x)}
\;\; ,
%		(12)
\end{eqnarray}
where $z = \sqrt{(1-y^2)/(1+x^2)} ~$. As for $S^{+-}_C(t)$, the
approximation $(|p/q|^2 + |q/p|^2)/2 \approx 1$ is rather safe
in the standard model. 

\vspace{0.4cm}

For the $B^0_d$-$\bar{B}^0_d$ system
we show the dependence of $S^{+-}_C(t)$ on
the decay time $t$ in Fig. 1, where $x \approx 0.723$ and
$y \approx 0$ \cite{PDG} (accordingly, $\phi_x \approx 0.626$ and
$\phi_y\approx 0$) have been taken.
We find that $S^{+-}_{-1}(t)$ and $S^{+-}_{+1}(t)$ become maximal 
at the positions $\Gamma t = (\pi +\phi_x)/x \approx 5.2$
and $\Gamma t = (\pi-\phi_x)/x \approx 3.5$, respectively.
The phase interval between these two line shapes, 
amounting to $2\phi_x/x$, also measures the rate of $B^0_d$-$\bar{B}^0_d$ 
mixing 
%%%%%%%%%%%%%
\footnote{The $C=+1$ $B^0_d\bar{B}^0_d$ events can in practice be
produced just above the $\Upsilon (4S)$ energy threshold, i.e., above
$M_B + M_{B^*}$ but below $2M_{B^*}$, whereby the $B^{*0}_d$ and
$\bar{B}^{*0}_d$ mesons decay radiatively, leaving 
$B^0_d\bar{B}^0_d\gamma$ with the $B^0_d\bar{B}^0_d$ pair in the
$C=+1$ state. In this case one has to pay for the cost that
the $b\bar{b}$ cross section above the $\Upsilon (4S)$ resonance
is smaller than that on the resonance \cite{BB}.}.
%%%%%%%%%%%%
%%%%%%%%%%%%%%%%%%
\begin{figure}
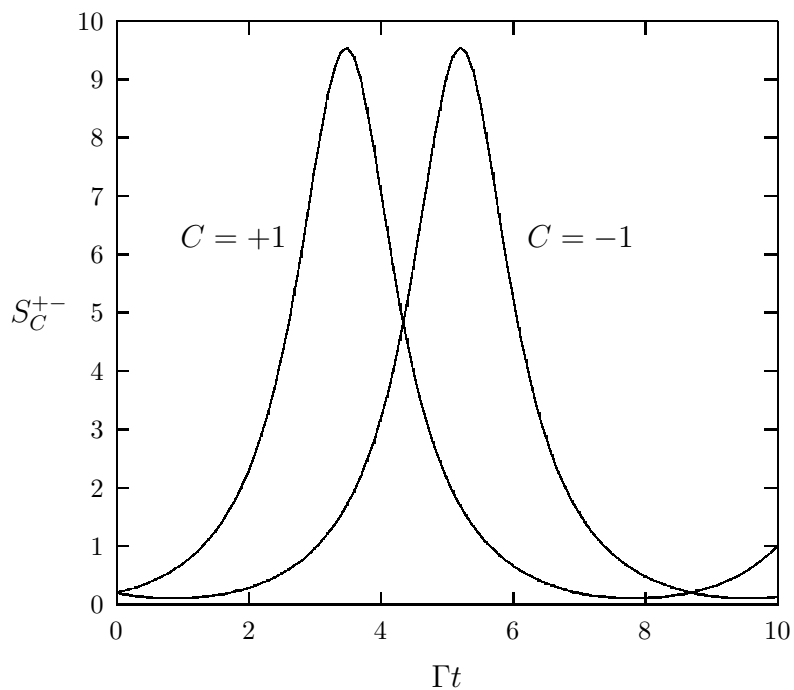

% GNUPLOT: LaTeX picture
\setlength{\unitlength}{0.240900pt}
\ifx\plotpoint\undefined\newsavebox{\plotpoint}\fi
\sbox{\plotpoint}{\rule[-0.200pt]{0.400pt}{0.400pt}}%

\caption{Ratios of the same-sign to opposite-sign dilepton
events changing with the decay time $t$ at the $\Upsilon (4S)$
resonance, where $x \approx 0.723$ and $y \approx 0$ 
have been taken.}
\end{figure}
%%%%%%%%%%%%%%%%%%

\vspace{0.4cm}

For the $D^0$-$\bar{D}^0$ system one has the following 
conservative bound on the mixing rate: $x<0.1$ and $y<0.1$
(satisfying $x^2+y^2<0.01$),
which were obtained from the wrong-sign semileptonic decays
of neutral $D$ mesons at the $90\%$ confidence level \cite{PDG,Dmixing}.
The relative magnitude of $x$ and $y$ remains unclear, as
the theoretical estimates involve too large uncertainty
due to the long-distance effects \cite{Burdman}.
In Fig. 2 we illustrate the time-dependent behavior of
$S^{+-}_C(t)$ with three types of inputs: (a) $x \approx y \approx 0.06$;
(b) $x \approx 0.08$ and $y \approx 0$; and (c) $x \approx 0$
and $y\approx 0.08$. 
We see that the line shape of $S^{+-}_C(t)$ for the $x\ll y$ case 
is clearly distinguishable, when $\Gamma t\geq 5$, from
that for the $x\gg y$ case. A delicate analysis even allows to
discern the relative magnitude of $x$ and $y$. This provides us
a new possibility, different from those proposed previously in
the literature \cite{Liu}, to measure the rate of $D^0$-$\bar{D}^0$ mixing
%%%%%%%%%%%
\footnote{For $\tau$-charm factories running at the $\Psi (4.14)$
resonance, the coherent $D^0\bar{D}^0$ events can be produced
through the transitions $\Psi (4.14)\rightarrow 
\gamma (D^0\bar{D}^0)_{C=+1}$ and
$\Psi (4.14) \rightarrow \pi^0 (D^0\bar{D}^0)_{C=-1}$. Note
that the $C=-1$ $D^0\bar{D}^0$ events can also be produced
from the decay of the $\Psi (3.77)$ resonance \cite{Charm}.}.
%%%%%%%%%%
%%%%%%%%%%%%%%%%%%%
\begin{figure}
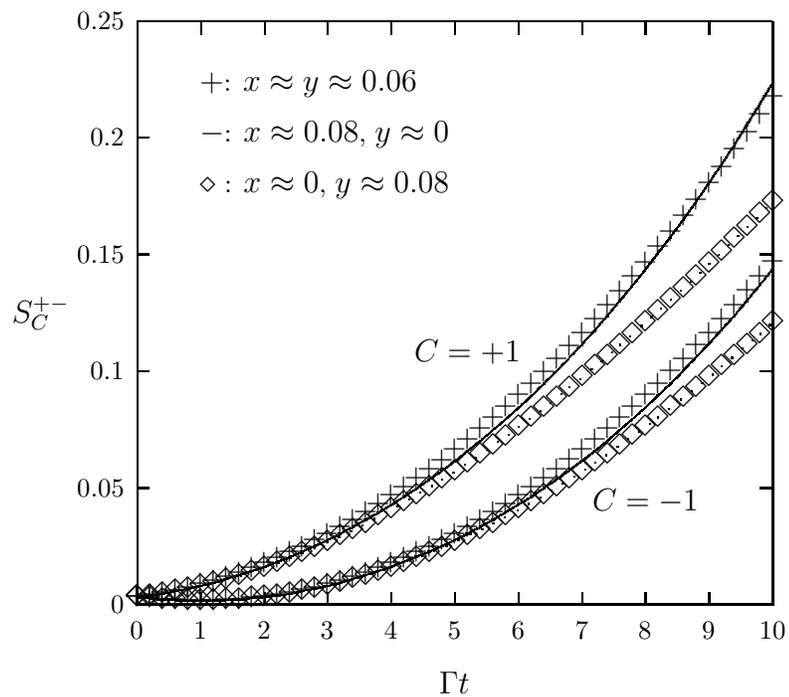

% GNUPLOT: LaTeX picture
\setlength{\unitlength}{0.240900pt}
\ifx\plotpoint\undefined\newsavebox{\plotpoint}\fi
\sbox{\plotpoint}{\rule[-0.200pt]{0.400pt}{0.400pt}}%

\caption{Illustrative plot for ratios of the same-sign
to opposite-sign dilepton events changing with the decay
time at the $\Psi (4.14)$ resonance.}
\end{figure}
%%%%%%%%%%%%%%%%%%%

\vspace{0.4cm}

For the $B^0_s$-$\bar{B}^0_s$ system we have $x>14$ from current
experimental data at the $95\%$ confidence level \cite{PDG}, and $y\sim 0.03$
from the latest theoretical calculation \cite{Beneke}. Hence the
behavior of $S^{+-}_C(t)$ depends mainly upon the value of $x$.
Taking $x\approx 20$ and $y\approx 0$ typically, one finds
that the oscillation term of $S^{+-}_C(t)$ is suppressed by a
factor $z\approx 1/x$. As a consequence $S^{+-}_C(t) \approx 1$
holds for variable values of $x$, i.e., the magnitude of
$S^{+-}_C(t)$ deviates little from unity. This property makes it
somehow difficult to determine the precise value of $x$ by 
measuring the time distribution of $S^{+-}_C(t)$ at the $\Upsilon (5S)$
resonance \cite{5S}.

\vspace{0.4cm}

{\Large\bf 4} ~
Now let us consider $CP$ violation in neutral $B$- or $D$-meson
decays into hadronic
$CP$ eigenstates at the $\Upsilon (4S)$ or $\Psi (4.14)$ resonance. 
In this case the semileptonic decay of one $P$ meson
serves to tag the flavor of the other $P$ meson decaying into
a nonleptonic $CP$ eigenstate. There are generally three 
different types of $CP$ asymmetries, arising from
$P^0$-$\bar{P}^0$ mixing itself, from the interference
between two decay amplitudes ({\it direct} $CP$ violation), and from the
interplay of decay and $P^0$-$\bar{P}^0$ mixing
({\it indirect} $CP$ violation). For the $B^0_d$-$\bar{B}^0_d$
system the typical magnitudes of
these three kinds of $CP$-violating effects are respectively
expected to be of ${\cal O}(10^{-3})$, ${\cal O}(10^{-2})$ to ${\cal O}(10^{-1})$,
and ${\cal O}(1)$ in the standard model. 
It is more difficult to classify 
the magnitudes of direct and indirect $CP$ asymmetries in
different decay channels of neutral $D$ or $B_s$ mesons, but
$CP$ violation in either $B^0_s$-$\bar{B}^0_s$ or 
$D^0$-$\bar{D}^0$ mixing is anticipated to be below
${\cal O}(10^{-3})$ within the standard model. Therefore
the neglect of tiny mixing-induced $CP$ violation, 
equivalent to taking $|q/p|\approx 1$ (as well as $y\approx 0$), is a
good approximation when we calculate the direct and indirect $CP$ 
asymmetries in most $B_d$, $B_s$ and $D$ decays. We obtain the time-dependent
decay rates as 
\begin{eqnarray}
R(l^{\pm}, f; t)_C & \propto &
|A_l|^2 |A_f|^2 \exp (-\Gamma t) \left [
\left (1 +|\lambda_f|^2 \right ) \pm \frac{1-|\lambda_f|^2}{\sqrt{1 +x^2}}
\cos (x\Gamma t + C\phi_x) \right . \nonumber \\
&  & \left . ~~~~~~~~~~~~~~~~~~~~~~~~~~~~~~~~~~~~~~~~~
\mp \frac{2 {\rm Im}\lambda_f}{\sqrt{1 +x^2}} 
\sin (x\Gamma t + C \phi_x) \right ] \; ,
%		(13)
\end{eqnarray}
where $f$ is the $CP$ eigenstate, and 
$\lambda_f = (q/p) \langle f|{\cal H}|\bar{P}^0\rangle/
\langle f|{\cal H}|P^0\rangle$ as defined before.
The $CP$ asymmetry is then given by
\begin{eqnarray}
{\cal A}^C_f(t) & = & \frac{R(l^-, f;t) - R(l^+, f; t)}
{R(l^-,f;t) + R(l^+, f;t)} \; \nonumber \\
& = & \frac{1}{\sqrt{1+x^2}} \left [
\frac{1-|\lambda_f|^2}{1+|\lambda_f|^2} 
\cos (x\Gamma t + C\phi_x) -
\frac{2 {\rm Im}\lambda_f}{1+|\lambda_f|^2}
\sin (x\Gamma t +C\phi_x) \right ] \; .
%		(14)
\end{eqnarray}
Clearly ${\cal A}^C_f(t)$ consists of both the direct $CP$
asymmetry ($|\lambda_f|\neq 1$) and the indirect one
(${\rm Im}\lambda_f \neq 0$). Measuring the time distribution
of ${\cal A}^C_f(t)$ can distinguish between these two sources
of $CP$ violation.

\vspace{0.4cm}

For illustration let us take the gold-plated channels
$B^0_d$ vs $\bar{B}^0_d\rightarrow J/\psi K_S$, which are
dominated by the tree-level quark transitions \cite{Xing93}, 
for example. It is well known that $|\lambda_{\psi K_S}|
\approx 1$ and 
${\rm Im}\lambda_{\psi K_S} = \sin (2\beta)$ hold,
where $\beta = \arg [-(V^*_{cb}V_{cd})/(V^*_{tb}V_{td})]$ is
an inner angle of the quark mixing unitarity triangle.
We are left with
\begin{equation}
{\cal A}^C_{\psi K_S}(t) \; =\; -\frac{\sin 2\beta}
{\sqrt{1+x^2}} \sin (x \Gamma t + C \phi_x) \; ,
%		(15)
\end{equation} to a high degree of accuracy. The behavior of
this $CP$ asymmetry changing with the decay time $t$ is illustrated in Fig. 3.
Certainly the weak phase $\beta$ can well be determined from
such a time-dependent measurement at the $\Upsilon (4S)$
resonance 
%%%%%%%%
\footnote{The result for the $C=-1$ case has been presented in 
Ref. \cite{Foland}, where the definition of $CP$ asymmetries is
different from ours in Eq. (14).}.
%%%%%%%%
%%%%%%%%%%%%%%%%
\begin{figure}
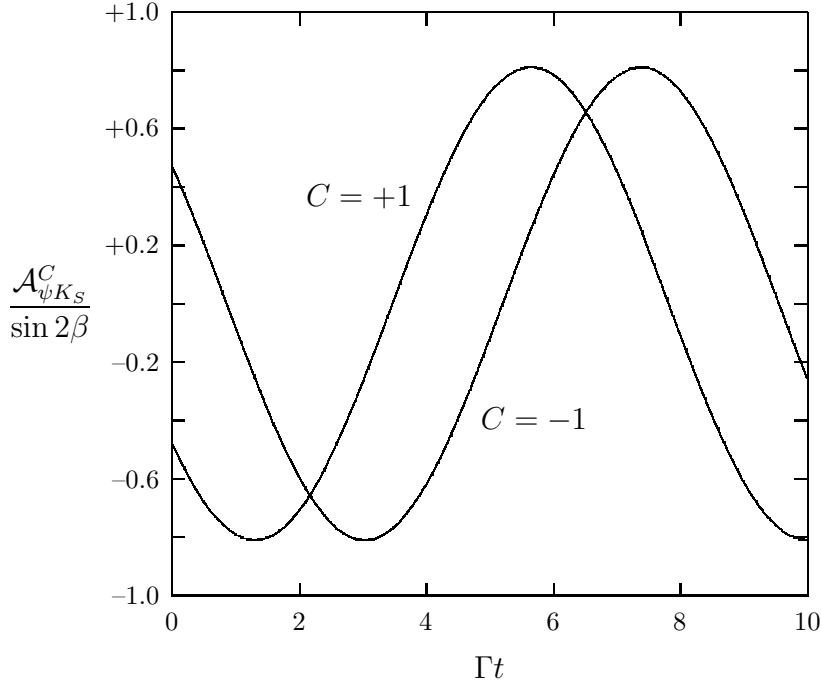

% GNUPLOT: LaTeX picture
\setlength{\unitlength}{0.240900pt}
\ifx\plotpoint\undefined\newsavebox{\plotpoint}\fi
\sbox{\plotpoint}{\rule[-0.200pt]{0.400pt}{0.400pt}}%

\caption{Time-dependent behavior of the $CP$ asymmetry in
$B^0_d$ vs $\bar{B}^0_d \rightarrow J/\psi K_S$ decays, 
where $x \approx 0.723$ has been taken.}
\end{figure}
%%%%%%%%%%%%%%%%

\vspace{0.4cm}

{\Large\bf 5} ~
Finally we consider the case that both $P^0$ and $\bar{P}^0$ 
mesons decay into a common non-$CP$ eigenstates. For neutral
$D$-meson decays, most of such decay modes occur through the quark
transitions $c\rightarrow s (u\bar{d})$ and 
$c\rightarrow d (u\bar{s})$ or their flavor-conjugate processes.
For $B_d$ and $B_s$ decays, most of such decay channels take place
through the quark transitions $b\rightarrow q (u\bar{c})$ and
$b\rightarrow q (c\bar{u})$ or their flavor-conjugate processes
(for $q=d$ or $s$). The typical examples of such decay channels include
$D^0$ vs $\bar{D}^0\rightarrow K^{\pm}\pi^{\mp}$,
$B^0_d$ vs $\bar{B}^0_d \rightarrow D^{\pm}\pi^{\mp}$, and
$B^0_s$ vs $\bar{B}^0_s \rightarrow D^{\pm}_s K^{\mp}$ decays
%%%%%%%%%%%%
\footnote{To extract the weak phase $\beta$ and 
$\beta'$ a study of $B_d$ and $B_s$ decays into the non-$CP$
eigenstates $D^{*\pm}D^{\mp}$ and $D^{*\pm}_sD^{\mp}_s$, in which the
penguin effects are negligibly small, is also
of particular interest \cite{Xing98}.}.
%%%%%%%%%%%

\vspace{0.4cm}

For simplicity we concentrate only on the decay modes in which no
direct $CP$ violation exists, i.e., the decay amplitudes of 
$P^0\rightarrow f$ and $\bar{P}^0\rightarrow \bar{f}$ are 
governed by a single weak phase. We also take $y \approx 0$,
as indirect $CP$ violation is primarily associated with the
mixing parameter $x$.
For coherent $P^0\bar{P}^0$
decays at the resonance, we make use of the semileptonic decay of one $P$ meson
to tag the flavor of the other $P$ meson decaying into $f$ or
$\bar{f}$. The time-dependent rates of such joint decay modes,
with the help of Eq. (8), are given as follows:
\begin{eqnarray}
R(l^-, f; t)_C & \propto &
|A_l|^2 |A_f|^2 \exp (-\Gamma t) \left [
\left (1 +|\lambda_f|^2 \right ) + \frac{1-|\lambda_f|^2}{\sqrt{1 +x^2}}
\cos (x\Gamma t + C\phi_x) \right . \nonumber \\
&  & \left . ~~~~~~~~~~~~~~~~~~~~~~~~~~~~~~~~~~~~~~~~~
- \frac{2 {\rm Im}\lambda_f}{\sqrt{1 +x^2}} 
\sin (x\Gamma t + C \phi_x) \right ] \; ,
\nonumber \\
R(l^+, \bar{f}; t)_C & \propto &
|A_l|^2 |A_f|^2 \exp (-\Gamma t) \left [
\left (1 +|\bar{\lambda}_{\bar f}|^2 \right ) + 
\frac{1-|\bar{\lambda}_{\bar f}|^2}{\sqrt{1 +x^2}}
\cos (x\Gamma t + C\phi_x) \right . \nonumber \\
&  & \left . ~~~~~~~~~~~~~~~~~~~~~~~~~~~~~~~~~~~~~~~~~
- \frac{2 {\rm Im}\bar{\lambda}_{\bar f}}{\sqrt{1 +x^2}} 
\sin (x\Gamma t + C \phi_x) \right ] \; ,
%		(16)
\end{eqnarray}
where $\bar{\lambda}_{\bar f} = (p/q) \langle \bar{f}|{\cal H}|P^0\rangle
/\langle \bar{f}|{\cal H}|\bar{P}^0\rangle$, and the relationship
$|\bar{\lambda}_{\bar f}| = |\lambda_f|$ holds.
The time-dependent $CP$ asymmetry turns out to be
\begin{eqnarray}
{\cal A}^C_{f\bar{f}}(t) & = & \frac{R(l^-, f;t) - R(l^+, \bar{f}; t)}
{R(l^-,f;t) + R(l^+, \bar{f};t)} \; \nonumber \\
& = & \frac{{\rm Im} \left ( \bar{\lambda}_{\bar f} - \lambda_f
\right ) \sin (x\Gamma t +C\phi_x)}
{\sqrt{1+x^2} \left (1 + |\lambda_f|^2 \right ) +
F(\lambda_f, \bar{\lambda}_{\bar f}, x\Gamma t + C\phi_x)} \; ,
%		(17)
\end{eqnarray}
in which $F$ is a function defined by
$F(z_1,z_2,z_3) = (1-|z_1|^2)\cos z_3 -{\rm Im}(z_1 +z_2)
\sin z_3$. Note that only the difference between
${\rm Im}\bar{\lambda}_{\bar f}$ and ${\rm Im}\lambda_f$,
which would vanish if the relevant weak phase were zero,
measures the $CP$ violation. 

\vspace{0.4cm}

Taking the decay modes $B^0_d$ vs $\bar{B}^0_d\rightarrow
D^{\pm}\pi^{\mp}$ for example, one finds that measuring
the $CP$ violating
quantity ${\rm Im} (\bar{\lambda}_{D^{\pm}\pi^{\mp}} -
\lambda_{D^{\mp}\pi^{\pm}})$ allows the determination of 
the weak phase $(2\beta +\gamma)$, where 
$\gamma = \arg [-(V^*_{ub}V_{ud})/(V^*_{cb}V_{cd})]$ is
another angle of the quark mixing unitarity triangle \cite{Xing95}.
This illustates that some attention is worth being paid to $CP$ 
violation in neutral $B$- and $D$-meson decays into hadronic
non-$CP$ eigenstates.

\vspace{0.4cm}

{\Large\bf 6} ~
In summary, we have derived the generic formulae for $P^0$-$\bar{P}^0$
mixing and $CP$ violation at the resonance where $P^0\bar{P}^0$
pairs can coherently be produced, for the case that only the decay-time
distribution of one $P$ meson is to be measured. Examples for
the $D^0$-$\bar{D}^0$, $B^0_d$-$\bar{B}^0_d$ and $B^0_s$-$\bar{B}^0_s$
systems are discussed. In particular, we point out a new 
possibility to measure $D^0$-$\bar{D}^0$ mixing in semileptonic 
$D$-meson decays at the $\Psi (4.14)$ resonance, and show that both
direct and indirect $CP$ asymmetries can be determined at the
$\Upsilon (4S)$ resonance with no need to order the decay times of
two $B_d$ mesons or to measure their difference. 

\vspace{0.4cm}

We expect that the formulae and examples presented here will be
useful for the physics being or to be studied 
at the $B$-meson and $\tau$-charm factories.

\end{document}